\documentclass[slac_one]{revtex4}
\usepackage{graphicx}
\usepackage{fancyhdr}
\usepackage{amsmath}
\def\bbar{\bar b}
\def\bb{ b\bbar}
\def\Jpsi{{J\mskip -3mu/\mskip -2mu\psi\mskip 2mu}}
\def\GeV{\text{GeV}\mskip -3mu/\mskip -2mu c\mskip 2mu}
\def\MeVc2{\text{MeV}\mskip -3mu/\mskip -2mu c^2\mskip 2mu}

\newcommand{\gra}{\mathrm{^o}}
\newcommand{\um}{\; \mathrm{\mu m}}
\newcommand{\mq}{\; \mathrm{m^2}}
\newcommand{\nbm}{\; \mathrm{nb^{-1}}}
\newcommand{\cmqs}{\; \mathrm{cm^{-2}s^{-1}}}
\newcommand{\etal}{\em{et al.}}
\begin{document}

\title{Status of LHCb} 

%

\author{Giovanni Carboni}
\affiliation{Dipartimento di Fisica,  
 Universit\`a degli Studi di Roma ``Tor Vergata'' \\ 
and Sezione INFN Tor Vergata, Roma, Italy}
\vspace*{2\baselineskip}
\author{\em On behalf of the LHCb collaboration} 

\begin{abstract}
The status of the LHCb experiment is presented. The experiment 
has been taking data since the LHC startup.  
The performances of the various sub-detectors are discussed 
and a preliminary measurement of the $\bb$ cross-section
is reported. The value is in agreement with expectations. 
\end{abstract}

\maketitle

\thispagestyle{fancy}


\section{INTRODUCTION}   
The LHC machine has a unique potential for the study of CP violation and rare
decays in the $b$ sector due to the combination of a large 
$\bb$ production cross-section and high luminosity.  
Even at an average 
luminosity $\mathcal{L}=2 \times 10^{32} \cmqs$, well below the 
machine design value,   
$10^{12} \; \bb$ pairs are produced per year, offering interesting 
perspectives in the quest for New Physics with an approach  
complementary to
those of ATLAS and CMS \cite{teubert}.
  
LHCb is the only experiment at LHC 
designed to exploit this potential with an optimized detector,
which incorporates precision vertexing and tracking systems,  
particle identification over a wide momentum spectrum 
and the capability to trigger down to very small 
transverse momenta. 
LHCb can work for years at constant luminosity, 
independently from the other intersections, thus  collecting 
clean events of constant quality.

The results presented here correspond to the machine startup phase,
with a reduced number of bunches and a value of $\sqrt{s} = 7$ TeV. 
The total integrated luminosity at the time of this conference  
(August 2010) amounts to  $1 \; \mathrm{pb^{-1}}$. Using the data collected
we show that the various subdetectors perform as planned. 
The $\bb$ 
production cross-section has been measured by two independent methods and
confirms the expectations on the sensitivity for the experiment. 

\section{DETECTOR PERFORMANCE}
\begin{figure}[htb]
\centering
\includegraphics[width=0.7\textwidth]{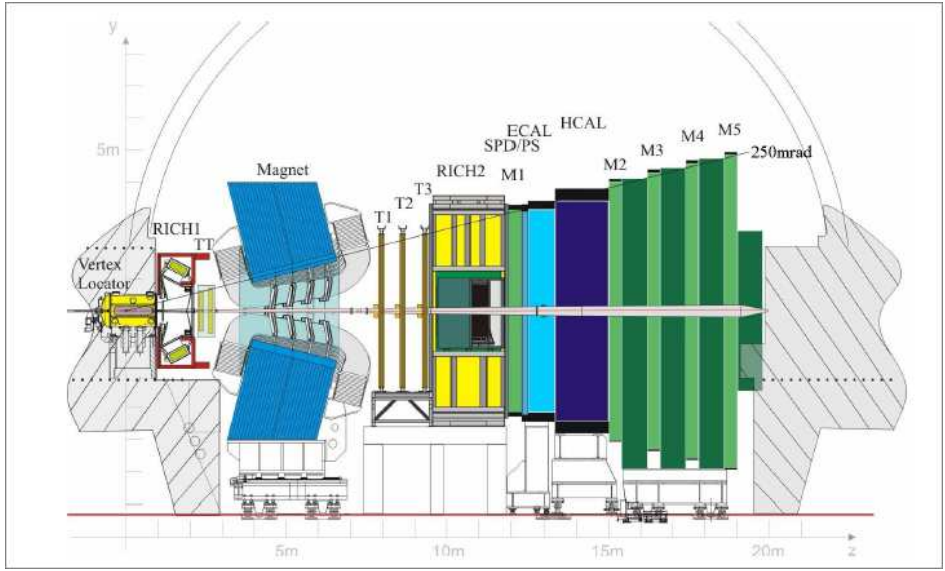}
\caption{Schematic side view of the LHCb detector. RICH1 and RICH2:
Ring Imaging Cherenkov; TT: Si Trigger Tracker;
T1-T3: downstream tracker stations; SPD/PS: Preshower; 
ECAL and HCAL: calorimeters; M1-M5: Muon Stations. See text
for additional details.}
\label{fig:detector}
\end{figure}

A schematic drawing of the LHCb detector is shown in Fig. \ref{fig:detector}. 
It is a 
single-arm open spectrometer using a large dipole magnet (bending power
3.6 Tm) for momentum measurement. 
This geometry exploits the fact the $\bb$ 
production is peaked in the forward direction. The apparatus is described in
detail in \cite{JINST}.

The LHCb detector is fully equipped since the startup of LHC, except for
the online trigger and DAQ farm which is still under completion. 
The geometry of LHCb is not well suited to rely on cosmic rays for calibration
purposes, therefore most of the calibration and alignment work has been
performed with particles from beam-beam collisions but also 
using beam-dump and beam-gas interactions. The performances 
of the various subdetectors are summarized below.  

\vspace*{\baselineskip}
\par \noindent
{\sffamily \bfseries \slshape Tracking.} The tracking 
system comprises a silicon strip detector (VErtex LOcator or VELO) to
reconstruct precisely the tracks close to the interaction region, followed by 
a larger silicon tracker before the magnet and by a hybrid tracker (silicon 
plus straw tubes) downstream of the magnet.  
The VELO is made of 21 silicon planes of semi-circular shape close to the
collision point, orthogonal to the beam direction (along the 
$z$ axis). The sensor strips are arranged to 
measure $r$ and $\phi$ with a $10 \gra - 20 \gra$ stereo angle to improve
spatial reconstruction. The VELO is mounted in a vacuum tank and the 
edges of the sensor planes are at few millimeters from the beam during data taking. They are retracted  to $\pm 2.9$ cm from the beam during machine injection.
The alignment among sensors is better than $5 \um$ and the fill-to-fill 
variations are less than this value. This ensures excellent stability and 
hit resolution. The cluster finding efficiency is 99.7 \% and the impact
parameter resolution for high-momentum tracks is $16 \um$ 
to be compared with 11.2 $\um$ from MC. Further improvement will be obtained 
by better alignment. The typical $z-$resolution for a primary vertex with 25 tracks is $\approx 90 \um$. From the analysis of 
$b \to \Jpsi$ decays a proper-time 
resolution of 60 fs has been obtained, giving excellent perspectives for 
time-dependent $B_s$ CP violation studies. 

\begin{figure}[htb]
\centering
\includegraphics[width=0.4\textwidth]{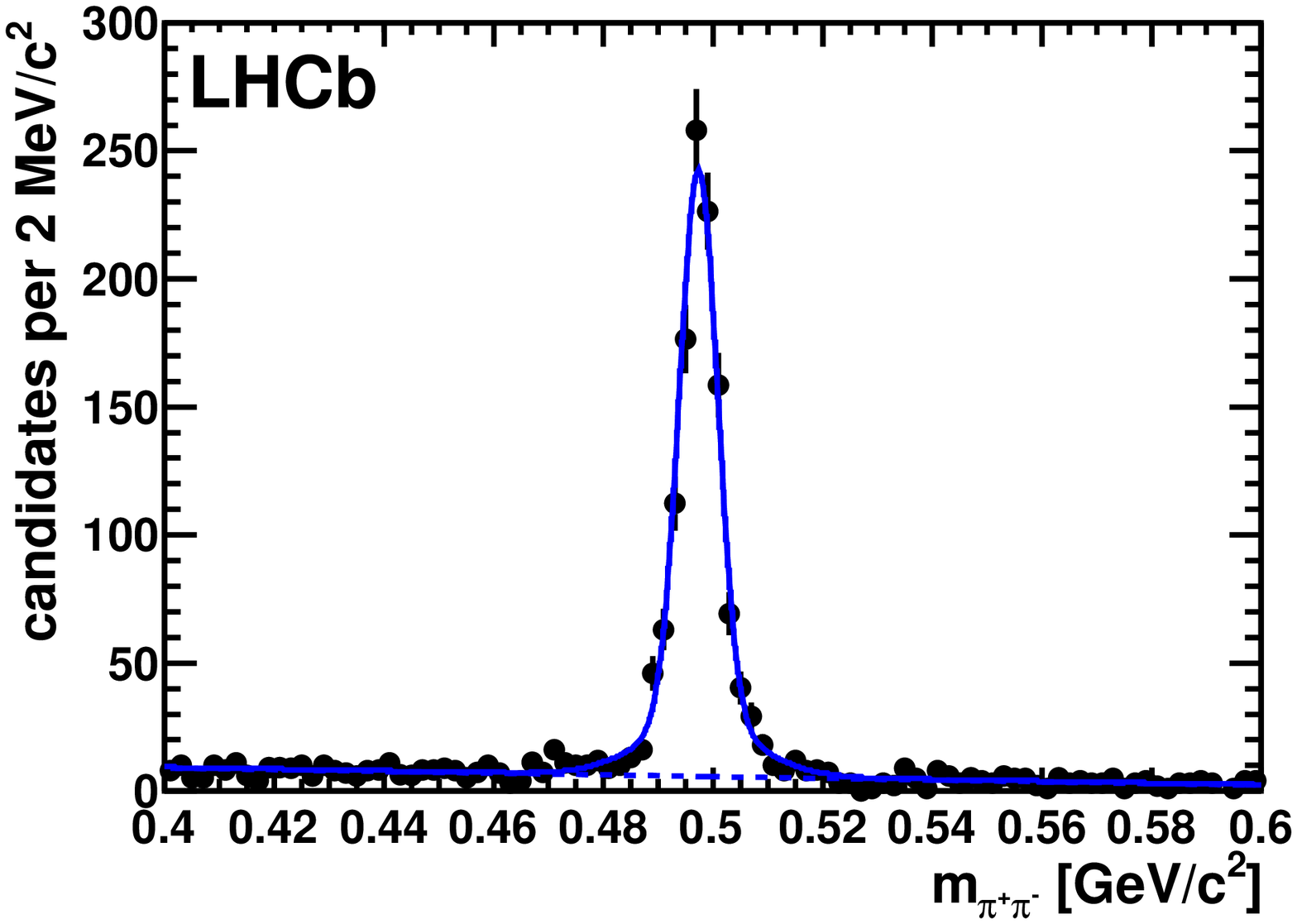}
\includegraphics[width=0.4\textwidth]{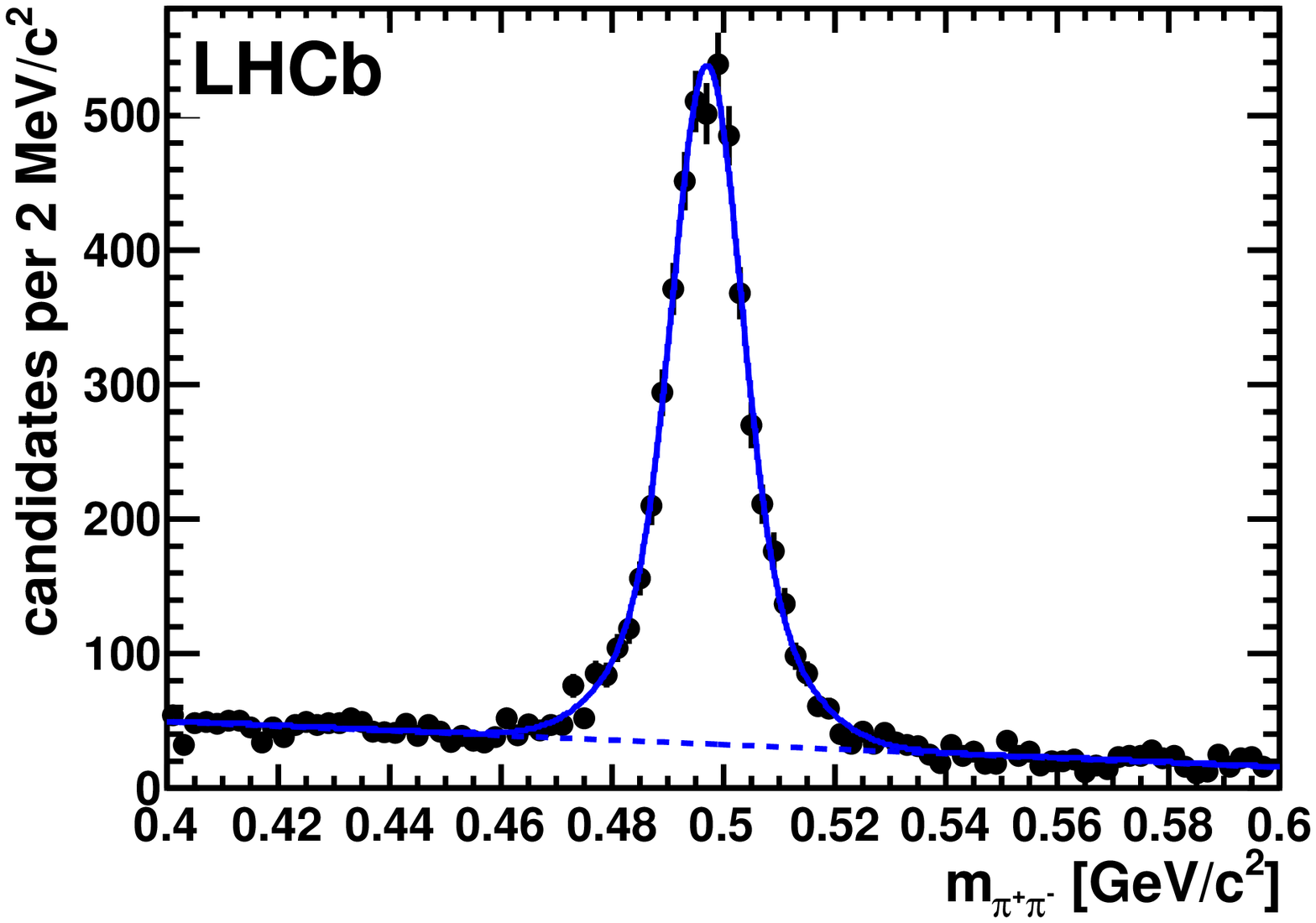}
\vspace*{-4mm}
\caption{Tracking performance. Left: mass peak for $K_\text{S}^0$ mesons 
with both pion tracks detected in the VELO. Mass resolution
is 5.5 $\MeVc2$ Right: same for $K_\text{S}^0$ 
decaying after the VELO, with mass resolution 9.2 $\MeVc2$.}
\label{fig:Ks}
\end{figure}

Following the VELO is  a large-area ($1.4 \times  2.1 \mq$) silicon strip 
detector (TT or Trigger Tracker) upstream of the magnet. Downstream of the
magnet there are three stations (T1 to T3) using silicon strips in the
inner part (IT) and straw tubes on the outer part (OT). The hit resolution 
is $\approx 55 \um$ in the TT and IT and $250 \um$ in the OT straw tubes.
Fig. \ref{fig:Ks} shows a typical $K_\text{S}^0$ mass peak as measured with or without
the use of the VELO. The mass resolutions are already very good, and better
alignment and calibration should allow the design values to be achieved,
which is about 30~\% smaller. 

\par
\begin{figure}[hbt]
\centering
\includegraphics[width=0.3\textwidth]{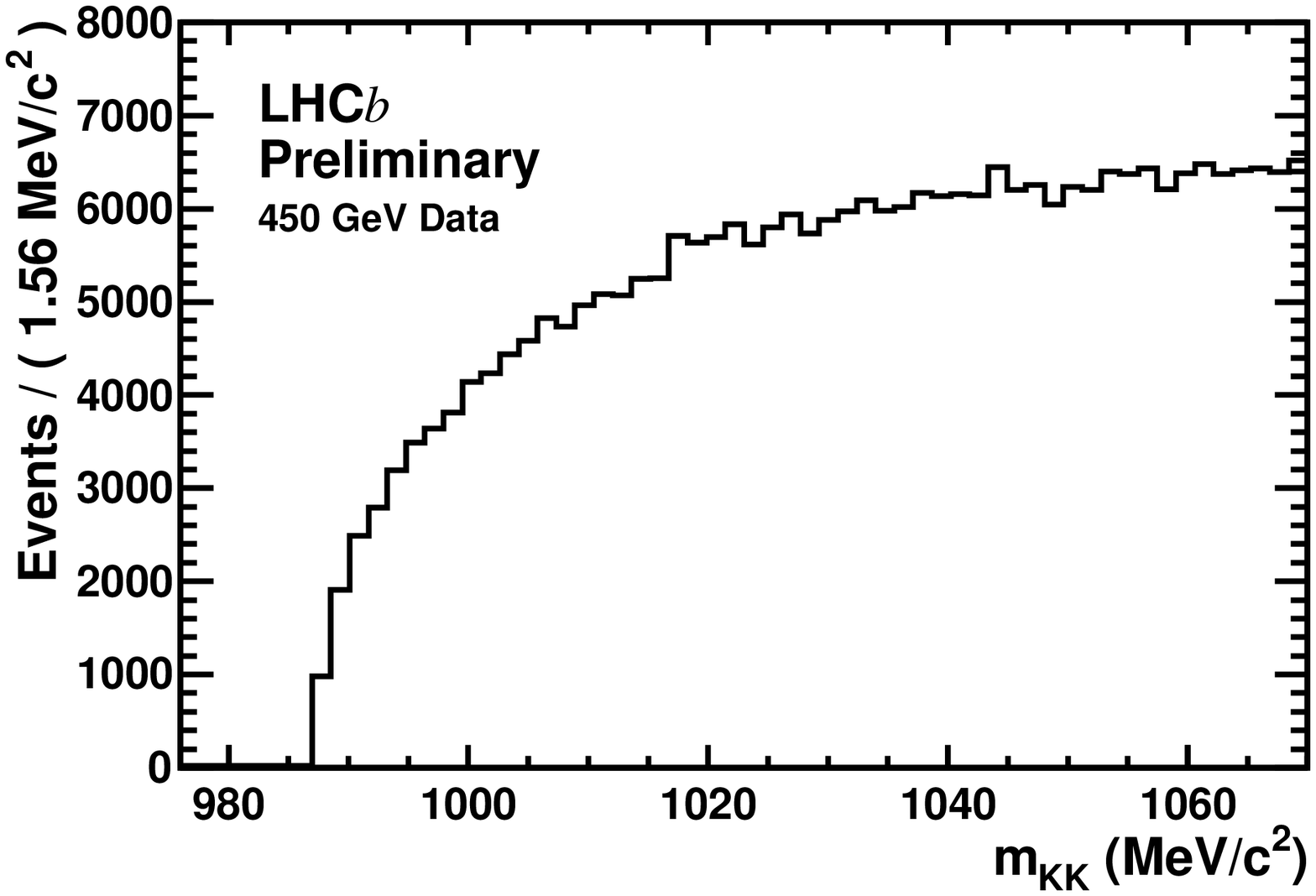}
\includegraphics[width=0.3\textwidth]{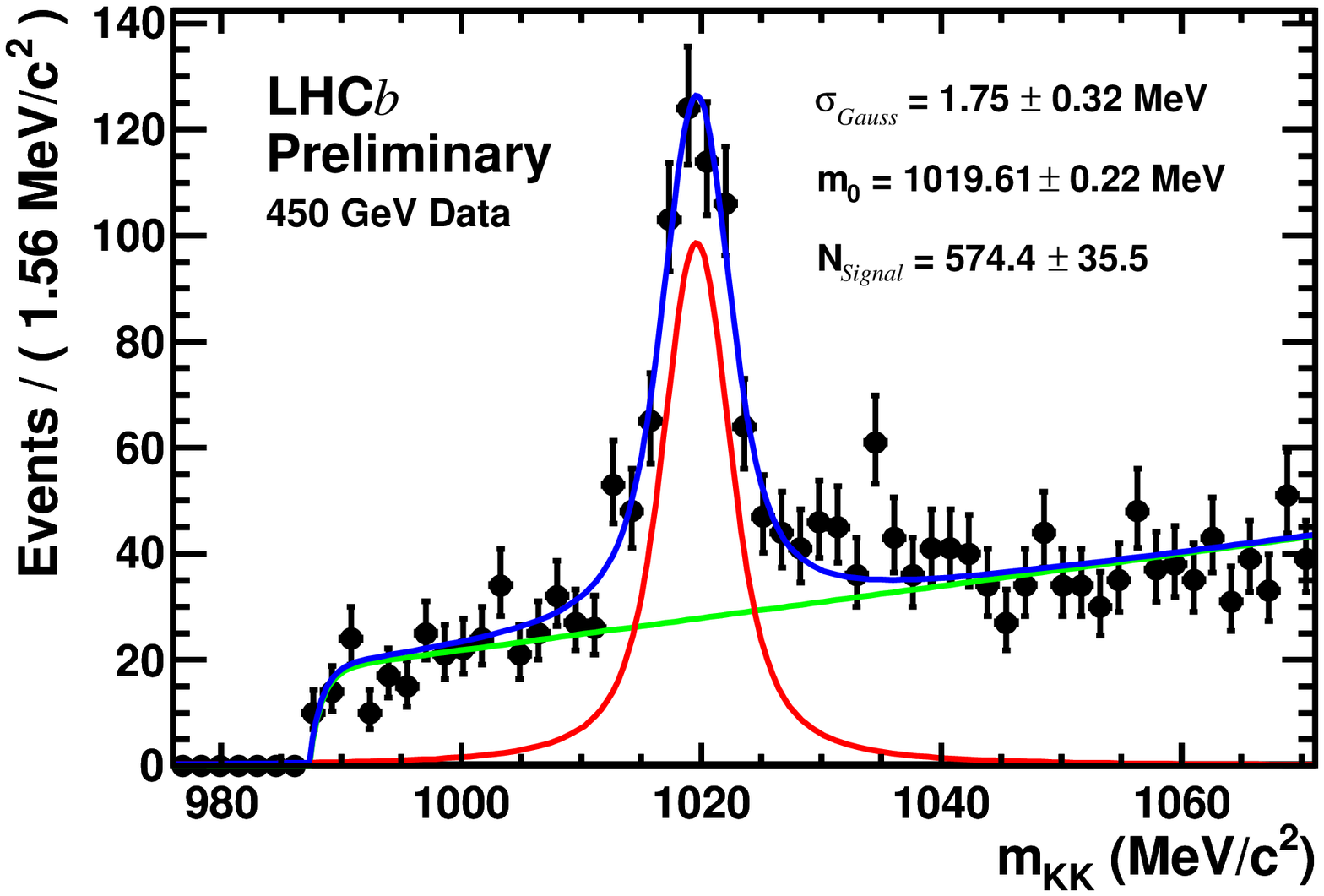}
\includegraphics[width=0.3\textwidth]{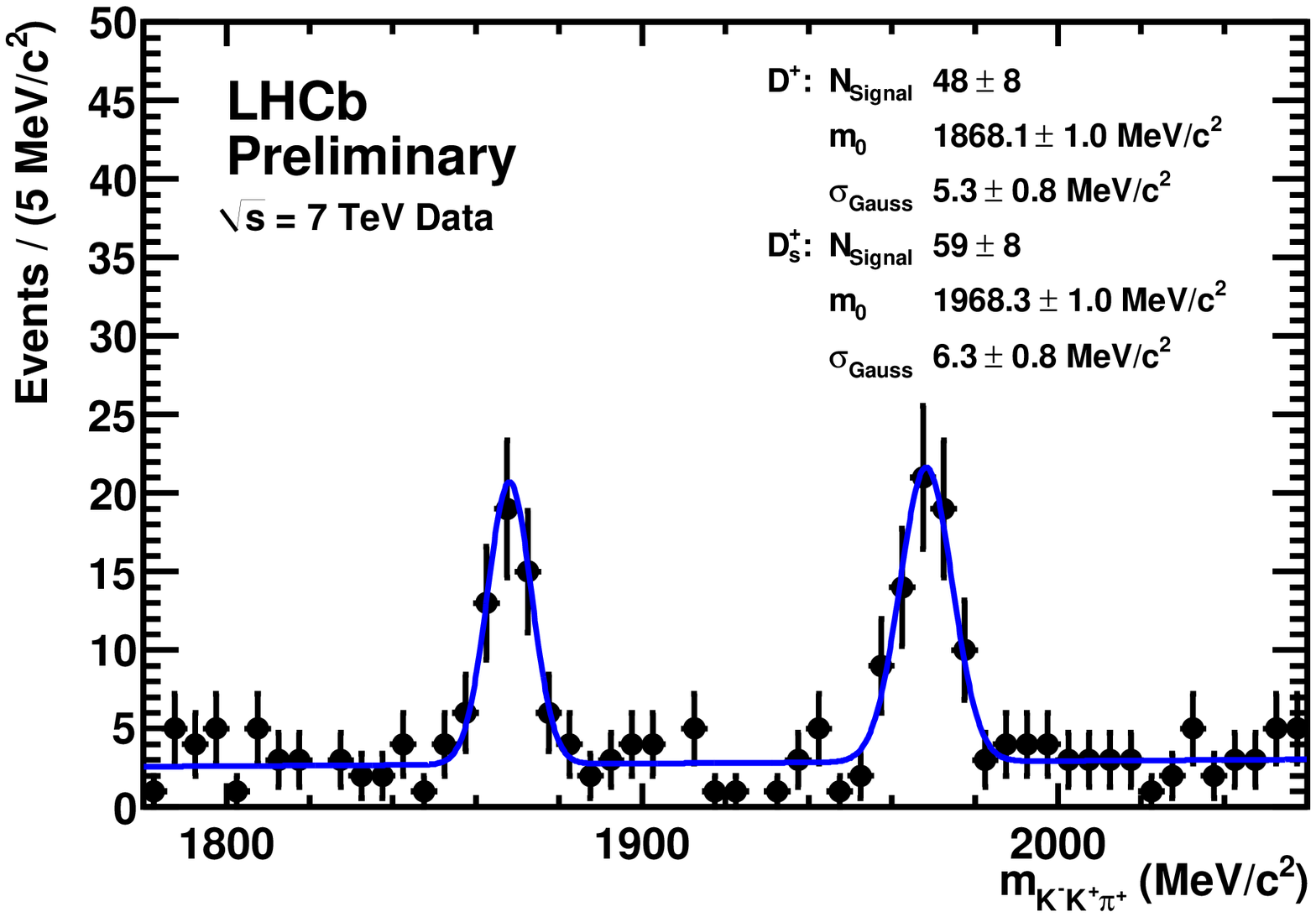}
\vspace*{-4mm}
\caption{RICH performance. Invariant mass distribution
for $\phi \to K^+ K^-$  
without  (left) and with 
 (center) $K$ identification. For $D^+$  and 
$D_s$ decaying to $K^+K^- \pi^+$ (right). }
\label{fig:phikk}
\end{figure}

 \noindent
{\sffamily \bfseries \slshape RICH.}
Two Ring Imaging Cherenkov detectors 
(RICH 1 and 2) with three radiators
are used to identify charged particles in the momentum range $2-100 \; \GeV$.
They play a  
crucial role for the clean reconstruction of several important channels. 
RICH1 uses
a silica Aerogel and a $C_4F_{10}$ gas radiator, 
whereas RICH2 uses $CF_4$ gas. 
The readout is via HPD pixel tubes. 
The angular resolution for the gas radiators 
($\sim 2.2$ and $\sim 0.9 \; \mathrm{mrad}$ respectively for RICH1 
and RICH2) is close to the design values. 
A resolution larger than expected is obtained for Aerogel, 
partially due to gas contamination of the radiator, and work is ongoing 
to calibrate every single tile of the aerogel wall. 
The 
average efficiency for $K^{\pm}$ detection is about 95~\% with a $\pi \to K$
misidentification probability of 7 \%. 
Fig. \ref{fig:phikk} shows the effect of the RICH detectors in 
reconstructing three benchmark channels.

\par \noindent
{\sffamily \bfseries \slshape Calorimetry and Muon Detection.}
Level-0 (L0) triggering and particle 
identification of electrons, photons and muons
are provided by dedicated subdetectors. The Hadron Calorimeter
(HCAL), based on an iron/scintillator tiles design, is important for 
triggering on multi-hadron decays. In a typical case like {\em e.g.} 
$D^* \to D^0 (K \pi) \pi$ the combined L0 and HLT (High level Trigger)
efficiency is $60 \pm 4 \%$, in good agreement with the Monte Carlo
prediction (66 \%).

\begin{figure}[htb]
\centering
\includegraphics[width=0.3\textwidth]{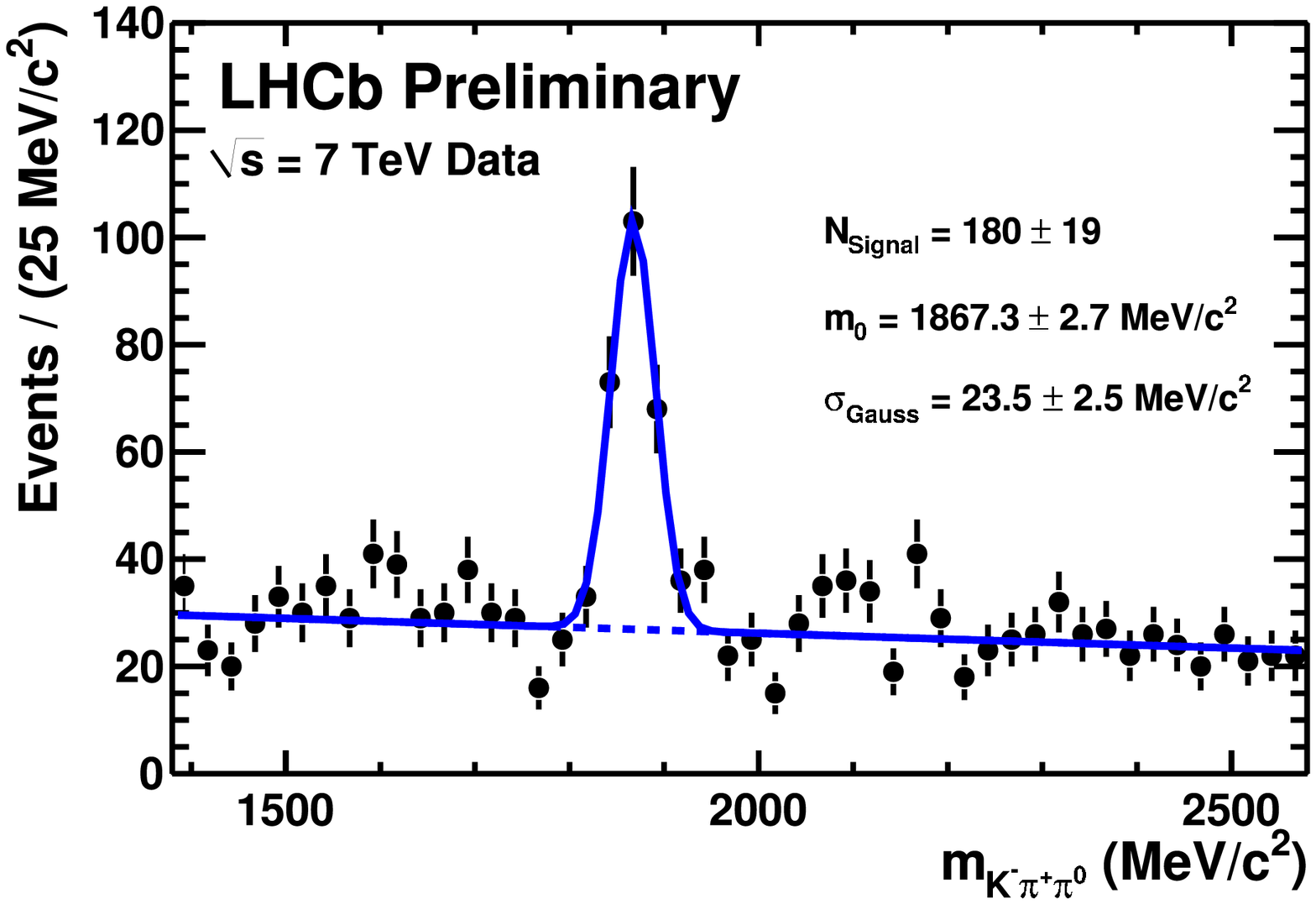}
\includegraphics[width=0.3\textwidth]{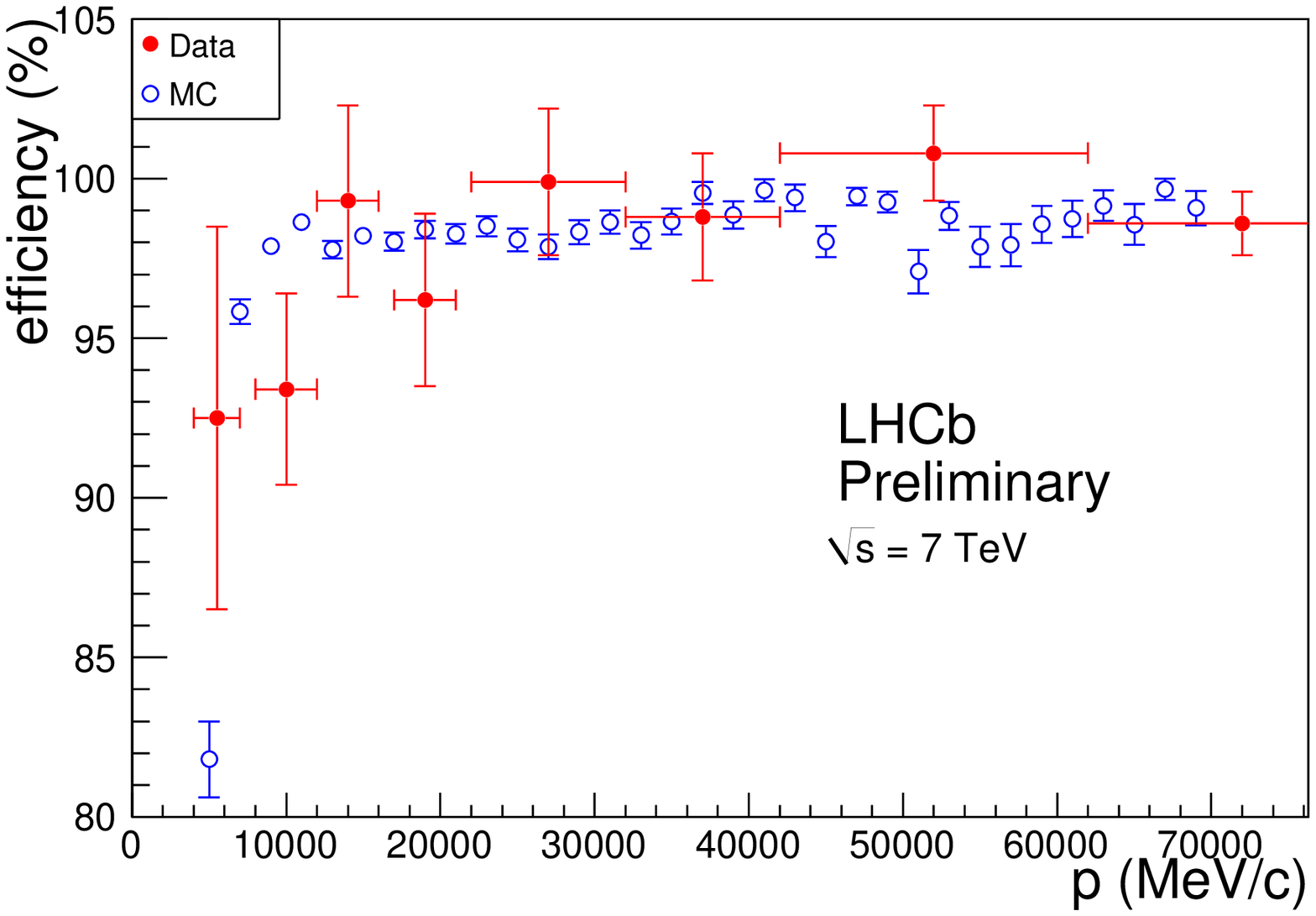}
\includegraphics[width=0.3\textwidth]{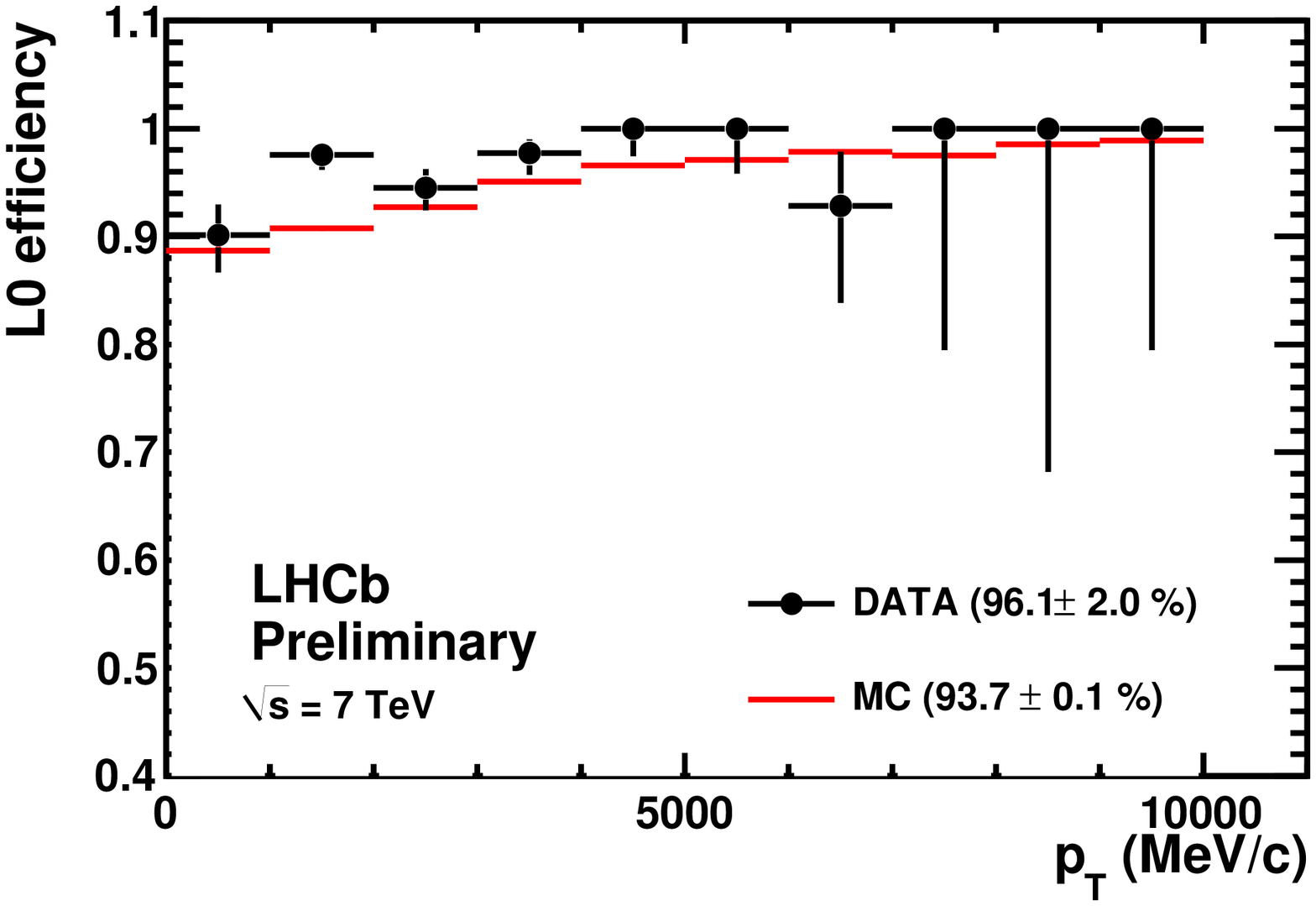}
\caption{Left:  $D^0 \to K^- \pi^+ \pi^0$, with $\pi^0$ 
reconstructed by ECAL and PS. Center: muon ID efficiency {\em vs.}
muon momentum. Right: L0 trigger efficiency for $\Jpsi \to \mu \mu$
{\em vs.} $\Jpsi$ $p_T$.}   
\label{fig:pi0}
\end{figure}

For electrons and photons 
the Electromagnetic calorimeter (ECAL) and the PS/SPD Preshower detectors
are used. The ECAL is a Shashlik-type sampling calorimeter, and its 
energy scale is calibrated to the 2 \% level, yielding a
resolution $\sigma_M = 7.2 \; \MeVc2$ on the $\pi^0$ mass. 
Reconstruction 
of multi-body decays with one $\pi^0$ in the final state is 
particularly encouraging  (Fig. \ref{fig:pi0}).

The MUON System uses five stations,M1-M5, 
equipped with 1368 fast MWPC chambers
supplemented by 24 Triple-GEM  chambers 
in the small-angle region of the first
Muon Station (M1). Since many interesting channels contain muons the
performance of the muon trigger and of muon identification are
particularly important. An average muon ID efficiency 
$\epsilon(\mu) = 97.3 \pm 1.2 \%$ has been measured on data, with
a misidentification probability for $\pi \to \mu$ around 2 \%
(see Fig. \ref{fig:pi0}). The same figure shows the L0 trigger efficiency for 
$\Jpsi \to \mu \mu$ events {\em vs.} $\Jpsi$ $p_T$. The average efficiency
measured on data ($96.1 \pm 2.0 \; \%$) agrees with  
the Monte Carlo prediction. 
 
\vspace*{\baselineskip}
\par \noindent
{\sffamily \bfseries \slshape Trigger architecture.}
The LHCb trigger is based on a powerful and very flexible 
multilevel architecture. Its task is extremely challenging
since  only $\sim 1/100$ of the events contain a $\bb$
and in addition the interesting channels have branching ratios
smaller than $10^{-3}$. 
At the lowest level (L0) the trigger 
is designed around dedicated electronics 
hardware operating synchronously with the
40 MHz bunch-crossing frequency 
and uses high-$p_T$ particles in the CALO 
and MUON systems to reduce the data rate from the 
40 MHz bunch crossing frequency to 1 MHz. The entire 
detector can be readout at this frequency. The output of
L0 is fed to the software High Level Trigger (HLT), running
asynchronously on a computer farm that will have 14000 processor
cores in its final phase. 
 The HLT adds the informations from the
tracking subdetectors (HLT1) and passes the interesting events 
for further analysis to HLT2. HLT2 has the capability 
of full event reconstruction to select the events to 
be written on tape at 2 kHz. In the 
present startup phase the demand on the trigger is very much 
reduced and therefore the HLT2 has been running in pass-through
mode, and the selections cuts for $p_T$ and impact parameter 
in L0 and HLT1 have been 
softened. The total efficiency for $B$ decays is at present 
$\simeq 70 \; \%$ for hadronic modes, and $> 90 \; \%$ for leptonic modes. 
In addition the relaxed cuts yield a large efficiency for charm
events.

\section{{\em b}-PRODUCTION CROSS-SECTION RESULTS}
The first data taken with the LHCb experiment at 
7 TeV centre-of-mass energy have allowed the measurement 
of the $b(\bbar)$ production cross-section in the 
LHCb acceptance. Beyond the intrinsic 
interest of this measurement,  
knowledge of the $b$ yield is also critical 
in ascertaining the sensitivity of the LHCb experiment
in flavour physics. 

The preliminary measurements of $\sigma_b$ presented in 
this paper correspond 
to an integrated luminosity $\int \mathcal{L} dt \approx~14\; \nbm$
and have been 
obtained by two independent methods. The first method \cite{Jpsi} 
uses the 
inclusive decay $b \to \Jpsi X$. The second method \cite{1009.2731} 
exploits the 
semileptonic channel $b \to D^0 \mu \nu X$, with $D^0 \to K^- \pi$
\setcounter{footnote}{2}
\renewcommand{\thefootnote}{\fnsymbol{footnote}}
The analysis is briefly presented below.  

\subsection{\boldmath $b \to \Jpsi X$} 
\begin{figure}[htb]
\centering
\includegraphics[width=0.4\textwidth]{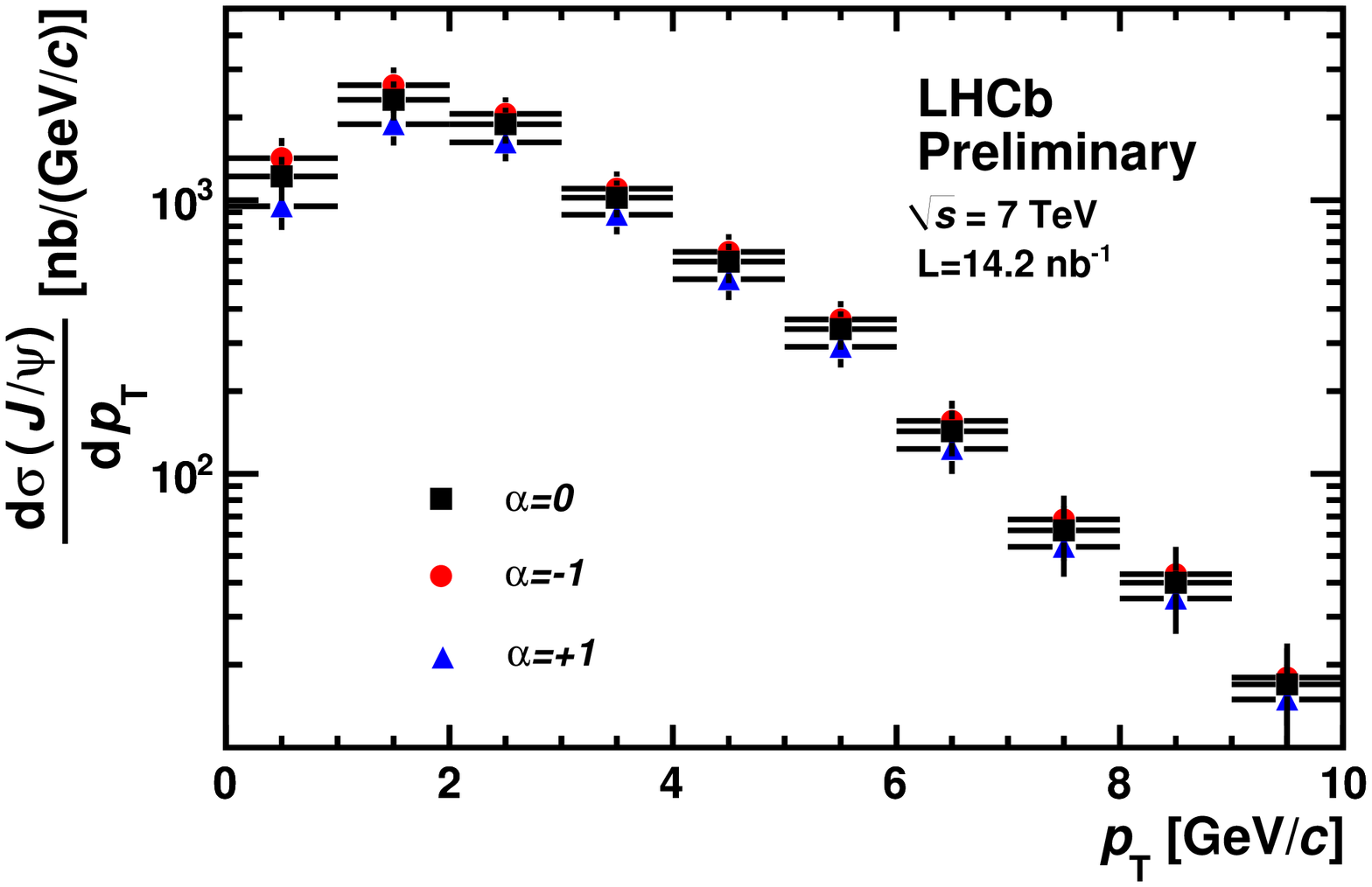}
\includegraphics[width=0.55\textwidth]{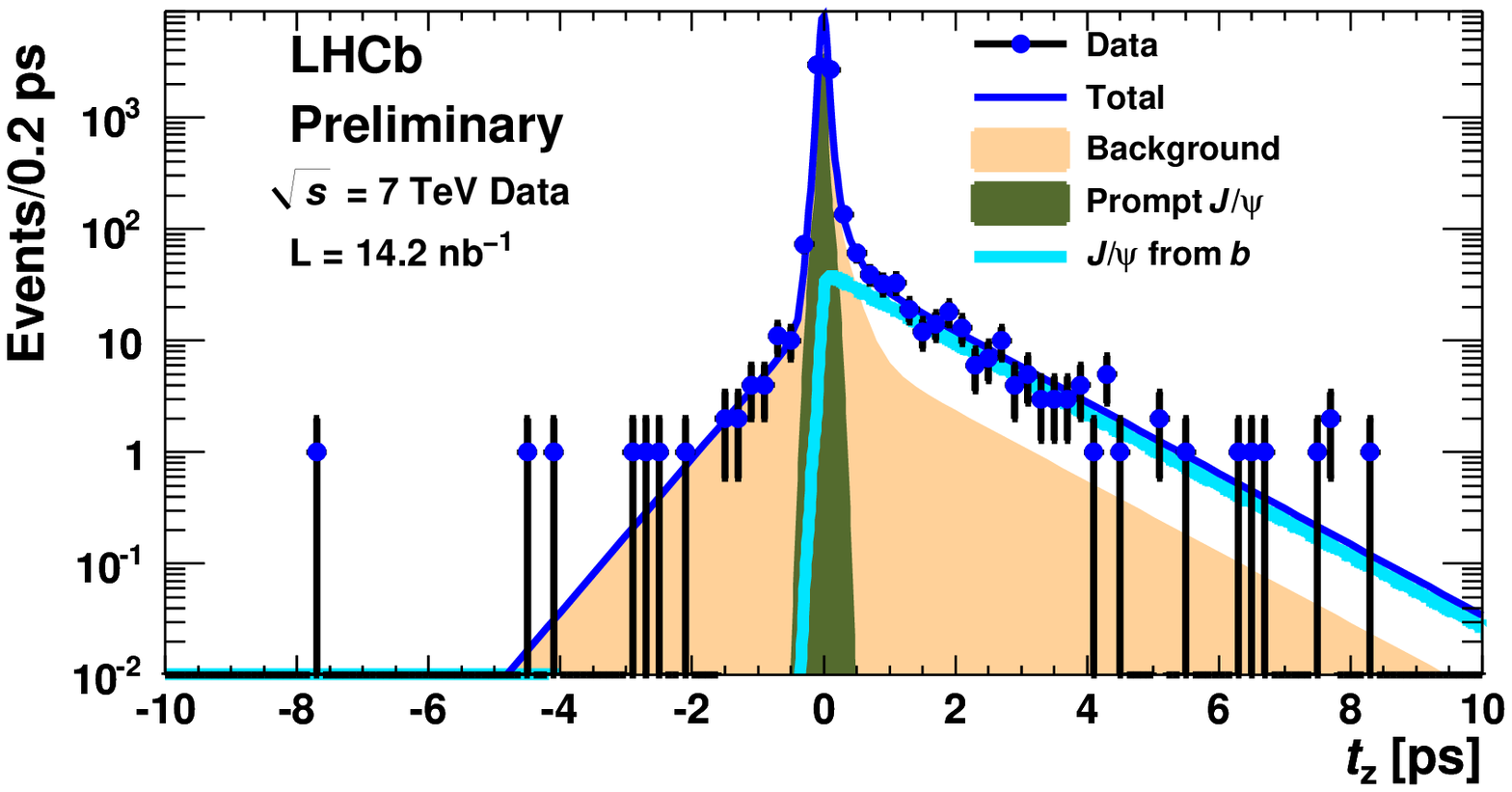}
\caption{Left: $d \sigma / dp_T$ for inclusive $\Jpsi$ production,
shown for three representative values of the $\Jpsi$ polarisation
(in the helicity frame). Right: pseudo proper time (see text) 
distribution of the $\Jpsi$ events, showing the various
contributions.}
\label{fig:sigmajpsi}
\end{figure}

For this study  the sample of $\Jpsi \to \mu \mu$ 
has been collected using the muon trigger, corresponding to an
integrated luminosity $\int \mathcal{L} dt =14.2 \; \nbm$. 
Full details of the analysis are presented in \cite{Jpsi}.
The trigger selected events with at least one muon track 
with $p_T > \;1.3 \;\GeV$ (HLT1). In the analysis 
selection cuts have been applied to the tracks and vertex quality. The
efficiency of the cuts and of the trigger and the 
acceptance of the detector have been evaluated using the LHCb Monte Carlo. 
The acceptance for the  $\Jpsi$ depends on its polarisation. The limited 
statistics does not allow yet for the measurement the polarisation parameter
$\alpha$, therefore the cross-section has a systematic uncertainty due to the 
unknown value of $\alpha$.  In 
Fig. \ref{fig:sigmajpsi} 
the  measured differential cross-section  
$d \sigma / dp_T$ is therefore presented for three representative values
of the polarisation parameter. The value integrated over the LHCb
acceptance is 
\begin{equation*}
\sigma(\text{inclusive } \Jpsi; p_T < 10 \; \GeV, 2.5 < y < 4) = 
7.65 \pm 0.19 \pm 1.10^{+0.87}_{-1.27} \; \mu \text{b} 
\end{equation*}
where the errors are, in the order,
statistical and systematical, the last error stemming from
the unknown polarisation. 

To separate the $\Jpsi$ produced in $b$ decays we use the pseudo proper
time $t_z = d_z M/P_z$ where $d_z$ is the projected distance bewteen
the primary and the dimuon vertx, and $M$ and $P_z$ are the $\Jpsi$
mass and longitudinal momentum respectively. A plot of the
$t_z$ distribution is shown in Fig. \ref{fig:sigmajpsi}. Fitting the
exponential decay term yields the fraction of $\Jpsi$ from 
$b$ decays, $f_b = 11.1 \pm 0.8 \; \%$ and a lifetime
$\tau_b = 1.35 \pm 0.10$ ps. From this the cross-section
for producing a $\Jpsi$ from $b$ decay is obtained:
\begin{equation*}
\sigma(\Jpsi \;\text{from }b; p_T < 10 \; \GeV, 2.5 < y < 4) = 
0.81 \pm 0.06 \pm 0.13 \; \mu \text{b} 
\end{equation*}
In this case, since many possible decay channels contribute
to the  $\Jpsi$ production,  an effective zero polarisation was assumed. By
extrapolating to the full solid angle using {\sc pythia} 6.4 \cite{pythia} 
the total $\bb$ production cross-section takes the value 
$\sigma(\bb) = 319 \pm 24 \pm 59 \; \mu \text{b}$. This value is 
in good agreement with the extrapolation from Tevatron energies
using {\sc pythia} with the colour-octet diagrams included. 
 
\subsection{\boldmath $b \to D^0 \mu \nu X$}
In this analysis the $D^0$ mesons from a parent $b$ are separated from those
directly produced on the basis of the impact parameter (IP) of the
reconstructed $D$ track measured with respect to the primary vertex.
The $D^0$ is reconstructed from the $K^-\pi^+$ decay. Full details of this
analysis are reported in \cite{1009.2731}. 

Two independent data samples, recorded at different times, were analyzed.
A so-called {\em microbias} sample comprised 
events with a very loose
trigger, requiring at least one track to be reconstructed in either the VELO 
or the tracking stations  ($\int \mathcal{L} dt
= 2.9 \nbm$). 
The second sample, referred to as {\em triggered}, 
used triggers designed to select a single muon 
($\int \mathcal {L} dt = 12.2 \nbm$).The two samples were 
analyzed independently and the results combined.  

\begin{figure}[hbt]
\centering
\includegraphics[width=0.7\textwidth]{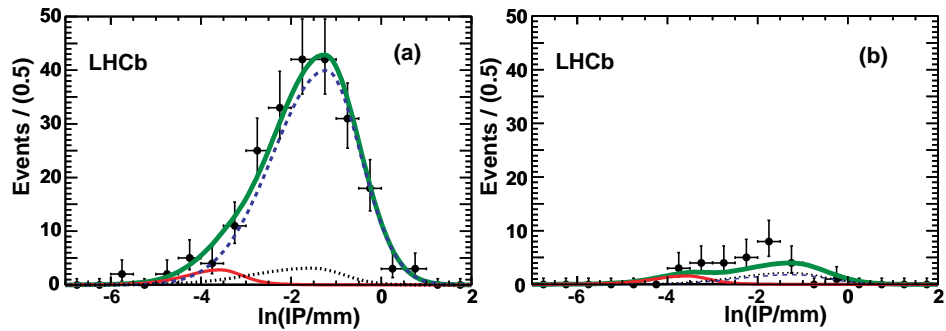}
\caption{\small Natural logarithm of the $D^0$ IP  in the 12.2 $\nbm$ triggered 
sample for (a) right-sign and (b) wrong-sign $D^0$-muon candidate combinations (see text).
 The dotted curves show the $D^0$ sideband backgrounds, the thin solid curves the Prompt yields, the
dashed curve the signal, and the thick solid curves the totals.
 } \label{fig:logip_bs}
\end{figure}

The $D^0\to K^-\pi^+$ candidates  
were matched with tracks identified as muons. Signal events from $b$ must have the same the same sign 
of the muon and of the kaon (RS events). The opposite associations (WS) indicated a background event.
The IP distributions of both RS and WS candidates, requiring that the
$K^-\pi^+$ invariant mass is within $20 \; \MeVc2$ of the $D^0$ mass, are shown in 
Fig.~\ref{fig:logip_bs} confirming that the WS events are highly suppressed. 

Since the $\theta$ angle of the $b$ hadron is obtained from the $D^0-\mu$ vertex, it is
possible to measure the differential cross-section $d \sigma_b/d \eta$. The results 
are reported in \cite{1009.2731} together with the predictions of two 
theoretical calculations (MCFM \cite{MCFM} and FONLL \cite{FONLL}). 
Integrating over the LHCb pseudorapidity acceptance yields the value
\begin{equation*}
\sigma_b(pp \to H_b X; 2 < \eta_b < 6 ) = 75.3 \pm 5.4 \pm 13 \; \mu \text{b}
\end{equation*}
Extrapolating to the full solid-angle using again {\sc pythia} 6.4 \cite{pythia} yields the value
$\sigma(\bb)~=~284 \pm 20 \pm 49 \; \mu \text{b}$, in good agreement with the
value obtained from the $b \to \Jpsi$ mode. 

\section{CONCLUSIONS}
The LHCb experiment is in excellent shape. More than 99 \% of the various 
sub-detectors are operational, and perform very close to the
expectations. Work is in progress to refine further 
the alignments and calibrations. The $b$-production cross-section has been measured in two different channels. Combining
the two measurements corresponds to a total $\bb$ production cross-section
\begin{equation*} 
\sigma(\bb; \sqrt{s} = 7 \text{ TeV}) = 298 \pm 15 \pm 43 \; \mu \text{b}
\qquad \text{(weighted average of two measurements)}
\end{equation*}
This value agrees with the expectations and confirms that the $b$ yield assumed 
in the design of LHCb was correct. With reasonable assumptions
on the integrated luminosity of the 2010-2011 run 
LHCb will be able to carry on its physics program and access signatures
sensitive to possible New Physics in CP violation and rare decays.


\begin{thebibliography}{99}

\bibitem{teubert} F. Teubert, these proceedings.


\bibitem{JINST}
A. Augusto Alves Jr. et al., (LHCb Collab.) 
``The LHCb Detector at the LHC'' 
JINST 3 (2008) S08005.

\bibitem{pythia}
T.~Sj\"{o}strand, S.~Mrenna and P.~Skands, ``PYTHIA 6.4: Physics and manual'', JHEP {\bf 05} (2006) 026.

\bibitem{Jpsi} 
LHCb Collaboration, Conference Note CERN-LHCb-CONF-2010-010 (2010).

\bibitem{1009.2731} LHCb Collaboration: E. Aaij {\etal}, arXiv:1009.2731v1 (2010), to be published 
in Physics Letters B.

\bibitem{MCFM}
The MCFM version 5.8 computer program was used to
evaluate the $b\overline{b}$ production cross-section. 
See J. M. Campbell and K. Ellis ``MCFM - Monte Carlo for FeMtobarn processes", at http://mcfm.fnal.gov/
and references therein. 

\bibitem{FONLL}
Private communication from M.~Cacciari, P. Nason, S. Frixione, M. Mangano, and G. Ridolfi.  See also
M.~Cacciari, S.~Frixione, M.~L.~Mangano, P.~Nason and G.~Ridolfi,
  JHEP { 0407} (2004) 33;
 M.~Cacciari, M.~Greco and P.~Nason,
  JHEP {\bf 9805} (1998) 007.


\end{thebibliography}
\end{document}